\documentclass[reprint,aps,ams,amsmath,prl,twocolumn,longbibliography,superscriptaddress]{revtex4-1}
\usepackage{graphics}
\usepackage{graphicx}
\usepackage{epstopdf,epsfig}
\usepackage{amsmath}
\usepackage{amssymb}
\usepackage{amsfonts}
\usepackage{mathrsfs}
\usepackage{bm}
\usepackage{color}
\usepackage{xcolor}
\usepackage{bbm}
\usepackage{hyperref}
\usepackage[normalem]{ulem}
\usepackage{comment}
\usepackage{soul}

\newcommand{\be}{\begin{equation}}
\newcommand{\ee}{\end{equation}}
\def\rr#1{(\ref{#1})}

\begin{document}

\title{Nonreciprocity of hydrodynamic electron transport in noncentrosymmetric conductors}

\author{E. Kirkinis}
\affiliation{Center for Computation and Theory of Soft Materials, Robert R. McCormick School of Engineering and Applied Science, Northwestern University, Evanston IL 60208 USA}

\author{L. Bonds}
\affiliation{Department of Physics, University of Washington, Seattle, Washington 98195, USA}

\author{A. Levchenko}
\affiliation{Department of Physics, University of Wisconsin-Madison, Madison, Wisconsin 53706, USA}

\author{A. V. Andreev}
\affiliation{Department of Physics, University of Washington, Seattle, Washington 98195, USA}

\date{\today}

\begin{abstract}
 We show that the nonreciprocity of hydrodynamic electron transport in noncentrosymmetric conductors with broken time-reversal symmetry (TRS) is significantly enhanced compared to the disorder-dominated regime. This enhancement is caused by the linear dependence of the viscosity of the electron liquid on the flow velocity, which is allowed in the absence of TRS and Galilean invariance. The resulting nonlinear flows break dynamical similarity and thus, in addition to the Reynolds number, they are characterized by a second dimensionless parameter, the nonreciprocity number. The latter is linear in velocity but independent of system size. 
 We determine the nonlinear conductance of a Hall bar and show that the nonreciprocal correction to the current can be of comparable magnitude to its reciprocal counterpart. 
 \end{abstract}

\maketitle

According to the Onsager reciprocity principle~\cite{Onsager-I,*Onsager-II}, the linear two-terminal conductance $G_0$ must be invariant under the time-reversal symmetry (TRS), which changes the
sign of the magnetic field
$\bm{B}$ and the magnetization of the system $\bm{M}$. In contrast, the nonlinear two-terminal  transport need not be reciprocal. 
In noncentrosymmetric conductors, i.e. nonpolar systems lacking inversion symmetry, due to the existence of an invariant that is linear in both the electric field $\bm{E}$ and the TRS breaking pseudovectors $\bm{B}$ or $\bm{M}$, the nonreciprocal contribution to the current density appears already in second order in the electric field $\bm{E}$~\cite{Sturman:1992,Rikken:PRL2001,Ivchenko:PRB2002}. Thus, in a two-terminal setup, the electric current $I$ through the system may be expressed as
\begin{equation}
 \label{eq:current_nonlinear}
    I = G_0 V +  G_2 V^2  , 
\end{equation}
where $V$ is the voltage bias, and the nonlinear part  of the conductance, $G_2$,  has a nonreciprocal (odd in the TRS-breaking perturbation) part. Nonreciprocity of nonlinear electron transport was the subject of extensive 
research in the context of mesoscopic systems in the disorder-dominated Drude regime~\cite{Spivak:PRL2004,Sanchez:PRL2004,Polianski:PRL2006,Tsvelik:PRL2006,Deyo:PRB2006}, with experiments on quantum dots, carbon nanotubes, and quantum wires~\cite{Linke:PRL2004,Cobden:PRL2005,Gossard:PRL2006,Marcus:PRL2006,Polianski:PRB2007}. In superconducting systems with broken TRS, nonreciprocity arises in both equilibrium current, i.e. the so called superconducting diode effect~\cite{Nadeem:2023}, and the dissipative regime~\cite{Liu:2024}.
For a recent review of nonreciprocal transport and optical phenomena in quantum materials see Ref.~\onlinecite{Yanase:2024}.

In this Letter, we develop the theory of nonreciprocal electron transport in noncentrosymmetric conductors with broken TRS in the hydrodynamic regime governed by momentum-conserving electron-electron scattering. The nondissipative effects of TRS breaking in the hydrodynamics of liquids, such as ferrofluids~\cite{Shliomis:1967,Odenbach:2002},  can be described by introducing nondissipative kinetic coefficients of the liquid~\cite{LL-V10}, which are odd under TRS as required by the Onsager symmetry. 
For instance, significant attention has been devoted to the study of odd, or Hall, viscosity of quantum liquids 
~\cite{Avron:1998,Haldane:2009,Read:2012,Son:2012,Messica:2025} and active matter systems~\cite{Banerjee:2017,Aggarwal2023, Kirkinis2023taylor,Kirkinis2024} with broken TRS.  We consider the effects of TRS breaking on the dissipative nonlinear hydrodynamic transport. 
In most high-mobility semiconductor heterostructures and graphene devices where the hydrodynamic electron transport has been demonstrated (see, e.g., reviews~\cite{Narozhny:2022,Fritz:2024} and references therein), the electron liquid lacks Galilean invariance. Recent experiments indicate that the electron liquid may spontaneously break TRS and inversion symmetry~\cite{Zhou:2021,Zhang:2024,LeoLi:arxiv2023,Chichinadze:2024}. 

We establish that nonlinear flows of nonreciprocal liquids
differ qualitatively from their
conventional counterparts in the following respect. Nonlinear hydrodynamic flows of conventional Newtonian liquids obey dynamical similarity~\cite{LL-V6,Batchelor1967}, 
which enables modeling of large-scale hydrodynamic phenomena in a lab; flows in homothetic systems can be mapped to one another by rescaling of the hydrodynamic velocity. All equivalent flows have the same Reynolds number $\mathcal{R} = U_0L/\nu$, where $U_0$ and $L$ are characteristic velocity and length scales, respectively, and $\nu$ is the kinematic viscosity. In contrast, we show that nonlinear flows of noncentrosymmetric electron liquids with broken time-reversal invariance do not possess dynamical similarity. Besides the Reynolds number, they must be characterized by an additional dimensionless parameter, which is proportional to $U_0$ and is odd under time-reversal.

The existence of the nonreciprocal component of $G_2$ in Eq.~\eqref{eq:current_nonlinear} requires breaking not only TRS but also a spatial symmetry, which distinguishes the flows in the forward and reverse directions. Similar to the situation in mesoscopic systems, this symmetry breaking may be associated with the device geometry~\cite{Andreev:PRL2006}, e.g. a flow in a funnel. Here we focus on the two-dimensional (2D) Poiseuille flow in the Hall bar,  which does not break the symmetry between the forward and reverse bias. In this case, the required spatial symmetry breaking arises in the noncentrosymmetric electron liquid itself. Below we study nonreciprocal transport for two different types of spatial symmetry breaking relevant to experiments:  \emph{i)} symmetry breaking of a vector type realized in 2D electron liquids with spin-orbit interaction in the presence of an in-plane Zeeman field, \emph{ii)} symmetry breaking of a tensor type realized  in valley-polarized graphene~\cite{Zhou:2021,Zhang:2024,LeoLi:arxiv2023,Chichinadze:2024}. The breaking of forward/reverse symmetry is described in \emph{i)} by a vector, which is linear in the Zeeman field, and in \emph{ii)} by an invariant rank-3 tensor that is allowed by the $C_3$ rotation symmetry in graphene. In the latter situation, the magnitude of the nonreciprocal conductance depends on the orientation of the Hall bar relative to the graphene lattice.

We show that nonreciprocity of dissipative transport is strongly enhanced in comparison to that in the Drude regime. The enhancement is caused by the linear dependence of the viscosity of the electron liquid on the hydrodynamic velocity $\bm{v}$. Such a dependence is allowed by symmetry
in noncentrosymmetric conductors with broken TRS. Its existence can be established in the framework of the Boltzmann equation for model systems of vector and tensor type symmetry breaking mentioned above. 
In a general situation, the liquid viscosity is described by a rank-4 tensor $\hat{\eta}$, which relates the viscous stress tensor 
\begin{equation}
\sigma_{ij}=\eta_{ijkl}V_{kl}, \quad V_{ij}=\frac{1}{2}\left(\partial_j v_i+\partial_i v_j\right), 
\end{equation}
to the strain rate tensor $V_{ij}$. For Galilean-invariant liquids, the viscosity tensor cannot depend on the flow velocity, and for isotropic incompressible liquids in two dimensions it can be reduced to the shear, $\eta$,  and odd, $\eta_H$, viscosities \cite{Avron:1998}. 
When both TRS and Galilean invariance are broken, a linear in the flow velocity correction to the viscous stress tensor is permitted, 
\begin{align}
    \label{eq:delta_sigma}
    \delta \sigma_{ij} = & \, T_{ijklm}v_k \partial_lv_{m}.
\end{align}

In vector-type symmetry breaking, where the TRS is broken by an external magnetic field $\bm{B}$, the rank-5 tensor $T_{ijklm}$ may be constructed using $\bm{B}$,  and the invariant tensors of Levi-Civita, $\epsilon_{ij}$, and Kronecker,   $\delta_{ij}$. This yields 
\begin{align}\label{eq:stress}
\delta \sigma_{ij}=&\eta\left[V_{ij}\left(\alpha\delta_{kl}B_lv_k+\widetilde{\alpha}\epsilon_{lk}B_kv_l\right)\right.\nonumber \\ 
&\left.+\beta v_iB_k\partial_kv_j+\widetilde{\beta} B_iv_k\partial_k v_j+\ldots\right].
\end{align}
Here we retained only several representative terms in the constitutive law and introduced phenomenological parameters $\alpha,\beta,\ldots$  whose values must be determined from a microscopic theory \footnote{In the framework of the Boltzmann transport equation, these parameters can be obtained following the method of Enskog (see e.g. \S 6 of Ref.~\cite{LL-V10}).
It is important that for non-parabolic electron dispersion $\epsilon(\bm{p})$, the locally-equilibrium distribution function in the presence of the hydrodynamic velocity $\bm{v}$,
$n_0 (\bm{p}) = \left[e^{ (\epsilon(\bm{p}) - \bm{p}\cdot \bm{v}-\mu)/T}+1 \right]^{-1}$, does not amount to a shift of momentum and the chemical potential $\mu$, but alters the shape of the Fermi surface. The distortion of the Fermi surface caused by $\bm{v}$ modifies all the kinetic coefficients of the electron liquid, including the viscosity. In electronic liquids with broken inversion and time-reversal symmetry, the modification of viscosity starts from the linear order in $\bm{v}$.}. 

In tensor-type symmetry breaking, which is realized in the recently discovered 
quarter-metal state in rhombohedral trilayer graphene~\cite{Zhou:2021}, the TRS breaking is caused by valley and spin polarization. 
Consequently, the system does not possess an in-plane vector breaking the TRS. However, due to the trigonal warping of the electron spectrum, the rotational symmetry of the electron liquid is lowered to $C_3$, which allows for an invariant rank-3 tensor $R_{ijk}$. The linear in $\bm{v}$ tensor $T_{ijklm}$ in Eq.~\eqref{eq:delta_sigma} can be constructed using $R_{ijk}$, $\delta_{ij}$, and $\epsilon_{ij}$.
It is more convenient to express this tensor relation in
the chiral complex coordinates $z=x+iy$ and $\bar{z}=x-iy$. In these coordinates,  a traceless stress tensor of an incompressible liquid is described by a single element $\delta \sigma_{zz} = \delta \sigma_{xx} -\delta \sigma_{yy} + 2 i \delta \sigma_{xy} $, and the invariant rank-3 tensor $R$ has only two non-vanishing components: $r \equiv R_{zzz}$ and $\bar{r} \equiv R_{\bar{z}\bar{z}\bar{z}}$~\cite{LL-V7} (since in Cartesian coordinates the tensor $R$ is real, $\bar{r}$ is the complex-conjugate of $r$). 
Upon a rotation by $2\pi/3$ the chiral coordinates are transformed as $z\to \epsilon z$, $\bar{z} \to \epsilon^{-1}\bar{z} $, where $\epsilon= e^{2\pi i/3}$. Therefore, we get 
\begin{align}
    \label{eq:stress_C3}
    \delta \sigma_{zz} = &  \,  \eta r V_{\bar{z}\bar{z}} v_z , \quad  \delta \sigma_{\bar{z}\bar{z}} =   \,  \eta \bar{r} V_{zz} v_{\bar{z}} , 
\end{align}
for the nonvanishing components of the nonreciprocal viscous stress tensor in Eq.~\eqref{eq:stress}.

For a given flow,
the ratio of the characteristic nonreciprocal stress in Eqs.~\rr{eq:stress} and \eqref{eq:stress_C3} to the standard viscous stress $(\sigma_0)_{ij} = 2 \eta V_{ij}$ defines a dimensionless parameter 
\be
\label{eq:nonreciprocity_number}
\mathcal{N}\sim \frac{\delta \sigma}{\sigma_0}\propto  U_0,
\ee
proportional to the characteristic flow velocity $U_0$ but  distinct from the Reynolds number. We refer to $\mathcal{N}$ as the \emph{nonreciprocity number}. 
It can be interpreted as the ratio of the typical flow velocity $U_0$ to the characteristic velocity scale associated with TRS breaking. In vector-type symmetry breaking the latter is proportional to the strength of spin-orbit coupling and the Zeeman field. In tensor-type, assuming that the degree of valley polarization is of order unity, the characteristic velocity is given by the Fermi velocity modulation caused by the trigonal warping. 
Since $\mathcal{N}$ is proportional to the flow velocity, but independent of system size, it breaks dynamical similarity of nonlinear flows. 

\begin{figure}[t!]
\centering
\begin{center}
\includegraphics[width=3in]{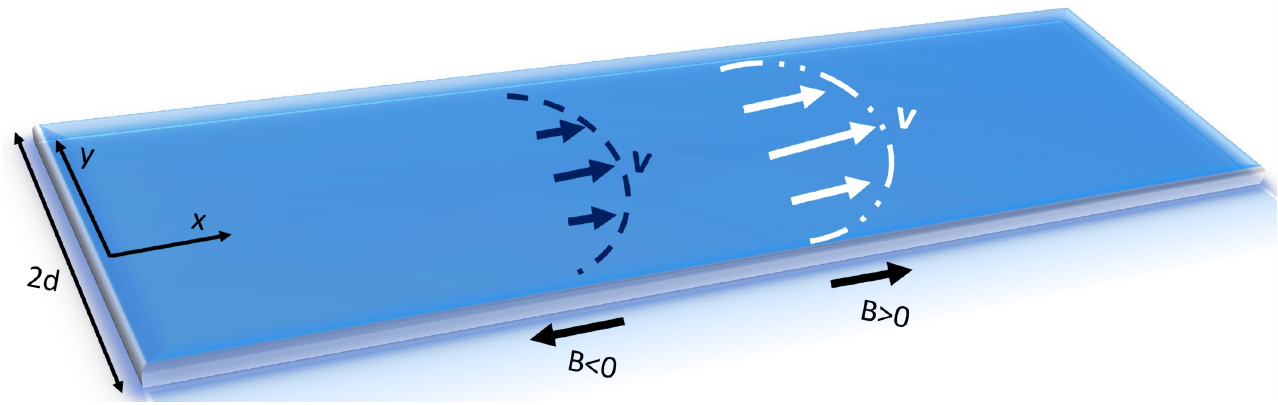}
\vspace{0pt}
\includegraphics[width=3.4in]{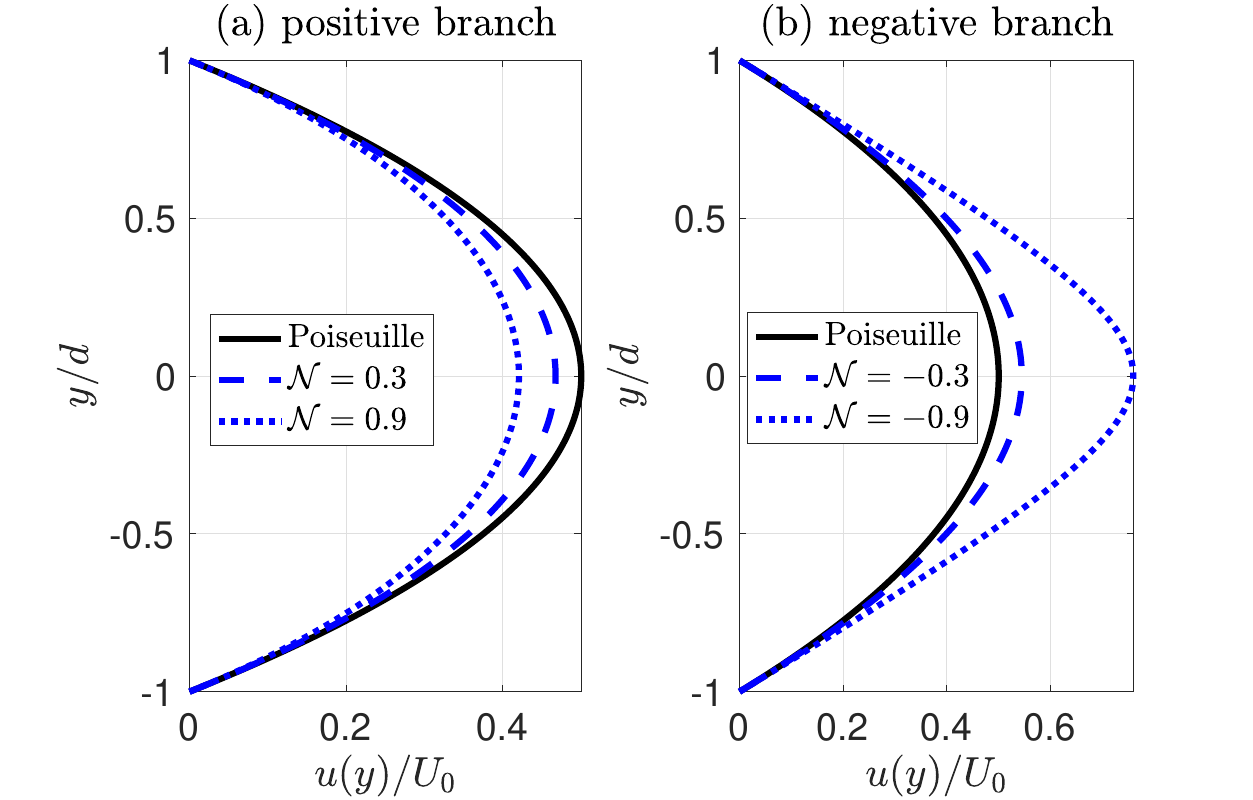}
\end{center}
\caption{2D nonreciprocal flow in a Hall bar of width $2d$. Nonreciprocal velocity profile $\bm{v}=u(y)\hat{\mathbf{x}}$ is controlled by the nonreciprocity number $\mathcal{N}$ in \rr{eq:nonreciprocity_number}, which depends on the orientation of the in-plane magnetic field $B$ in vector-type symmetry breaking, and on the valley polarization and the orientation of the graphene lattice in tensor-type. Panel (a) displays the velocity profile \rr{eq:u-flow} for  $\mathcal{N}=0.3$ and $0.9$, while panel (b) displays the velocity profiles for the time-reversed values $\mathcal{N}=-0.3$ and $-0.9$. In both cases the Poiseuille profile is restored for $\mathcal{N} \rightarrow 0$.}
\label{fig:Hall-bar}
\end{figure}

To obtain quantitative results for nonreciprocal electron transport in the hydrodynamic regime, we consider the experimentally relevant two-dimensional Poiseuille flow in the  Hall bar geometry shown in Fig. \ref{fig:Hall-bar}. 
Transport of momentum in a steady-state hydrodynamic flow is described by the force balance equation \cite{LL-V6}
\begin{equation} \label{NS}
n e E_i = \partial_j\Pi_{ij}, 
\end{equation}
where $n$ is the density of electrons, $e$ is the electron charge, $\bm{E}$ is the electric field, and $\Pi_{ij}$ is  the momentum flux tensor density. The latter is conventionally expressed in terms of the pressure $P\delta_{ij}$ and the viscous stress tensor $\sigma_{ij}$, as
\begin{equation}
\label{eq:Pi_ij}
\Pi_{ij}=P\delta_{ij}+\rho v_iv_j-2\eta V_{ij} -\delta\sigma_{ij}.
\end{equation}
In the absence of Galilean invariance, the expression $P \delta_{ij} + \rho v_i v_j$ for the equilibrium part of the momentum flux tensor in  Eq.~\eqref{eq:Pi_ij}  should be viewed
as a formal expansion of momentum flux density in an equilibrium state of uniform flow to second order in the powers of the flow velocity $\bm{v}$ with higher-order terms omitted. 

For a current flowing through a channel of width $2d$, as displayed in Fig.~\ref{fig:Hall-bar}, the velocity field has only one nonzero component $\bm{v} = u(y)\hat{\mathbf{x}}$, which depends only on the transverse coordinate $y$. The viscous force density is fully described by the single component of the viscous stress, $\sigma_{xy} (y)$. Its nonreciprocal part $\delta \sigma_{xy} = \eta N u \partial_y u$ is characterized by a single parameter $N$, which has units of inverse velocity. For a characteristic flow velocity $U_0$, the nonreciprocity number is then defined as $\mathcal{N} = N U_0$, and Eq.~\eqref{NS} reduces to
\begin{equation} \label{ns1}
\frac{en\mathcal{E}}{U_0} + \partial_y\left[ \eta\left(1+ \mathcal{N} \frac{u}{U_0}\right)\partial_y\frac{u}{U_0}\right] = 0, 
\end{equation}
where we absorbed the pressure gradient $\partial_xP$ into the electromotive force (EMF)  density $n e \mathcal{E}=neE-\partial_x P$. The latter is related to the voltage drop, $\mathcal{E} = V/L$, with $L$ being the length of the Hall bar. For Poiseuille flow, the convective nonlinear terms in the Navier–Stokes equation do not affect the velocity profile~\cite{LL-V6}. Therefore, the nonlinear two-terminal conductance arises only from the velocity dependence of the viscosity, and is entirely nonreciprocal.

For vector- and tensor- type of symmetry breaking the value of 
the nonreciprocity parameter $\mathcal{N}$ in 
Eq. \eqref{ns1}  is  determined from Eqs.~\eqref{eq:stress} and \eqref{eq:stress_C3}. For vector-type the precise form of $\mathcal{N}$ depends on the mutual orientation of the field with respect to the flow velocity. For tensor-type we have $\mathcal{N}=(r+\bar{r})U_0/2$. Notice that a rotation of crystalline axes by an angle $\theta$ relative to the flow changes $r \to r e^{3 i \theta}$. Therefore, the combination $r+\bar{r}$ in the nonreciprocity parameter exhibits a periodic modulation $\propto \cos(3[\theta - \theta_0])$. This is consistent with the three-fold rotation symmetry observed in nonreciprocity measurements in valley-polarized multilayer graphene \cite{LeoLi:arxiv2023,Chichinadze:2024}. 

Equation \eqref{ns1} can be readily integrated with the integration constants fixed by the boundary condition imposed on the flow. For simplicity, we apply no-slip conditions $u(\pm d)=0$. The more general case of a flow with a finite slip length does not alter the essential physics underlying nonreciprocity. Choosing $U_0$ as the Poiseuille flow velocity at the center of the channel, $U_0=en\mathcal{E}d^2/\eta$, we express the profile of the flow in the form
\begin{equation}\label{eq:u-flow}
u(y)=\frac{U_0}{\mathcal{N}}\left[\sqrt{1+\mathcal{N}\left(1-\frac{y^2}{d^2}\right)}-1\right].
\end{equation}
Equation \eqref{eq:u-flow} is valid for both positive and negative values of $\mathcal{N}$. At $\mathcal{N}\to 0$ it reproduces the Poiseuille flow profile, see Fig. \ref{fig:Hall-bar} for illustration. 

From the flow profile, we can compute the total current~\footnote{In systems lacking Galilean invariance, the total current density consists of the usual convective term, $enu$, and an additional contribution arising from the finite intrinsic conductivity~\cite{LAL:2022}. Assuming the particle density is sufficiently high, we neglect the latter.}. The resulting expression for the total current ($B>0$) takes a relatively simple form 
\begin{equation} \label{eq:I}
I =\int_{-d}^{+d}\!\! enu(y) dy =I_0 f(\mathcal{N}), \quad I_0=\frac{2}{3}\frac{(en)^2\mathcal{E}d^3}{\eta}, 
\end{equation}
where the dimensionless function is given by 
\begin{equation} \label{f}
f(\mathcal{N})=\frac{3}{2\mathcal{N}^{\frac{3}{2}}}\left[(1+\mathcal{N})\arcsin\sqrt{\frac{\mathcal{N}}{1+\mathcal{N}}}-\sqrt{\mathcal{N}}\right].
\end{equation}
The field-independent linear conductance, $G_0=(I/V)_{\mathcal{N}\to0}$, takes the value 
\begin{equation}\label{G0}
G_0=\frac{2e^2}{3}(nd^2)\frac{n}{\eta}\frac{d}{L}.
\end{equation}
It is inversely proportional to the viscosity of the electron fluid, which is the manifestation of the Gurzhi effect \cite{Gurzhi:1968}. This behavior was confirmed experimentally \cite{Kumar:2017,Brar:2023,Zeng:2024}. 

\begin{figure}[t!]
\centering
\includegraphics[width=3.2in]{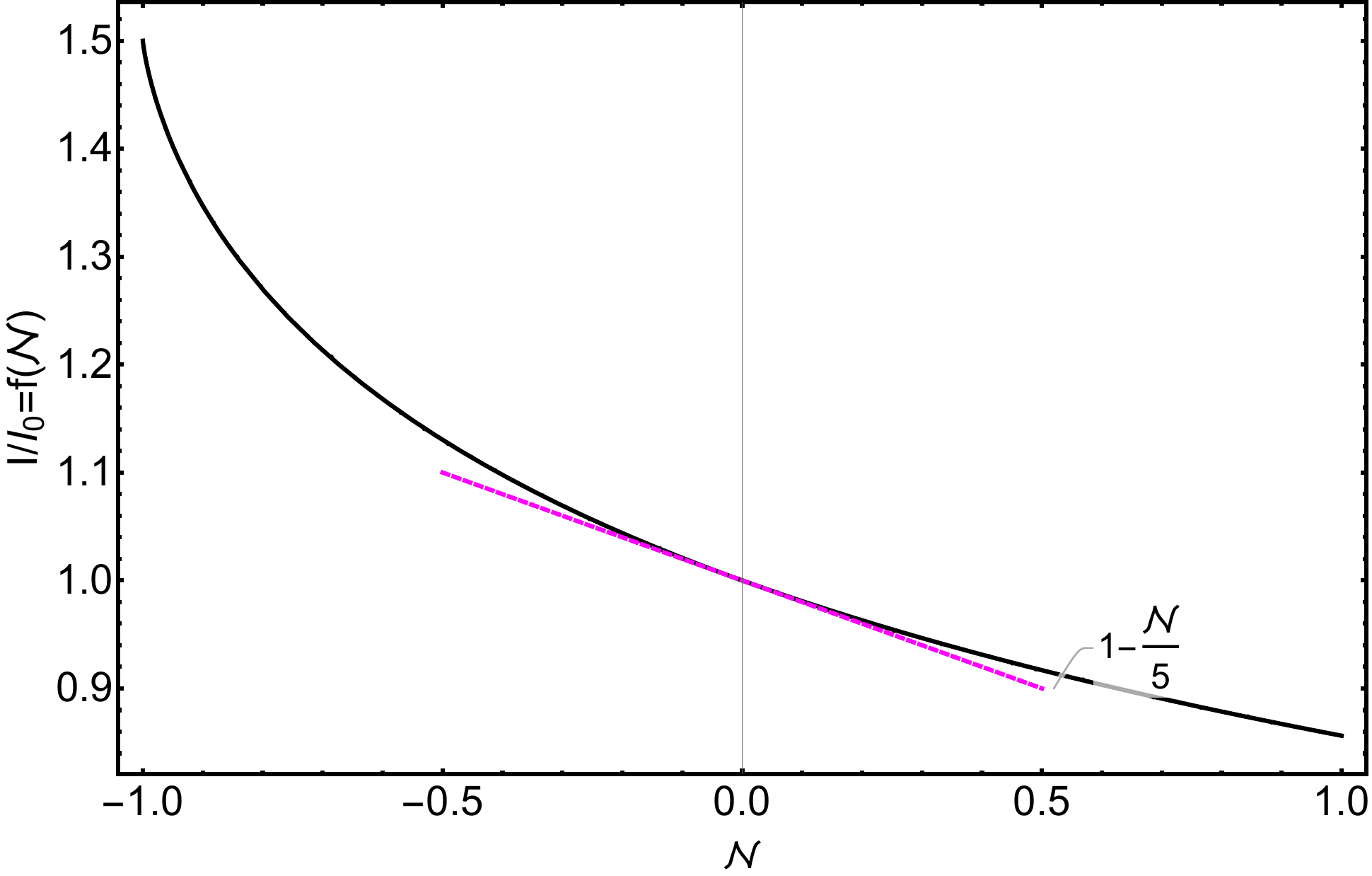}
\caption{Plot of the dimensionless current $I/I_0 = f(\mathcal{N})$ in Eq. \rr{eq:I} where $f$ is given by Eq. \rr{f}, vs. the nonreciprocity number for positive branch ($\mathcal{N}>0$) and negative branch ($\mathcal{N}<0$) of the flow profile, cf. Fig. \ref{fig:Hall-bar}. The dashed line denotes its linear approximation, $I/I_0 \approx 1-\mathcal{N}/5$, whose nonreciprocal correction to the conductance was calculated in Eq.~\rr{G2}.
At fixed bias, this shows the dependence of the nonlinear conductance on the TRS breaking strength.}
\label{fig:nlconductance}
\end{figure}

The applicability of Eqs.~\eqref{eq:I} and \eqref{f} at increasing bias is limited by the condition of stability of the laminar flow. A different condition, $\mathcal{N} \lesssim 1$, arises from expanding the viscous stress tensor to linear order in the velocity. 
Working to leading order in $\mathcal{N}$, we extract the nonlinear nonreciprocal correction to conductance of the system. In the notations of Eq. \eqref{eq:current_nonlinear} it takes the form
\begin{equation}\label{G2}
G_2=-\frac{G_0}{5}\frac{nd^2}{\eta L}\left\{\begin{array}{cc} (e\alpha B), & \text{vector-type}, \\ e|r|\cos(3[\theta-\theta_0]), & \text{tensor-type} .\end{array}\right.
\end{equation}
This is obtained by expanding the expression
$I/I_0 = f(\mathcal{N})\approx 1-\mathcal{N}/5$ in \rr{eq:I}, where $f$ is given by Eq. \rr{f}, and keeping only leading order terms in the nonreciprocity number $\mathcal{N}$. 
For TRS breaking caused by an in-plane magnetic field $\bm{B}$, the nonreciprocity number is $\mathcal{N}=\alpha BU_0$, where $\alpha$ depends on the spin-orbit coupling and the angle between $\bm{B}$ and the flow direction. 
In Fig. \ref{fig:nlconductance} we display the function $f(\mathcal{N})$ vs. $\mathcal{N}$, along with the limiting cases of weak and strong nonreciprocity.   

We now compare the magnitude of nonreciprocity in the hydrodynamic regime to that in the disorder-dominated regime. In the latter case, the ratio of the nonreciprocal to reciprocal current can be expressed as 
\begin{equation}\label{eq:dI-I-dis}
\frac{\delta I_{\text{dis}}}{I_0}\sim \varsigma\frac{e\mathcal{E}\ell_{\text{ei}}}{\epsilon_{\text{F}}},
\end{equation}
where $\ell_{\text{ei}}=v_{\text{F}}\tau_{\text{ei}}$ is the disorder-limited electron mean-free path, and $\varsigma$ is a dimensionless strength of TRS breaking. For example, for carbon nanotubes considered in Ref.~\onlinecite{Spivak:PRL2004}, $\varsigma\sim\Phi/\Phi_0$, where $\Phi$ is the magnetic flux through the nanotube and $\Phi_0=hc/e$ is the flux quantum. In the hydrodynamic regime, the corresponding estimate, for the relative magnitude of the nonreciprocal current is significantly enhanced in comparison to its disorder-dominated counterpart. In valley-polarized graphene, the degree of nonreciprocity of the electron spectrum  depends on the energy scale $\Delta$ associated with the trigonal warping. The corresponding  dimensionless  parameter $\varsigma\sim\Delta/\epsilon_{\text{F}}$, may in principle reach values of order unity. For the case of 2D electron liquid with Rashba spin-orbit coupling, this parameter can be estimated as a ratio of the Zeeman energy to the spin-orbit splitting. Considering the expression $U_0\sim\frac{d^2}{\eta}en\mathcal{E}$ for the typical flow velocity and estimating the shear viscosity in the Fermi liquid regime as $\eta\sim np_{\text{F}}\ell_{\text{ee}}$, where $\ell_{\text{ee}}=v_{\text{F}}\tau_{\text{ee}}$ is the inelastic mean free path, we obtain   
\begin{equation}\label{eq:dI-I-hydro}
\frac{\delta I_{\text{hydro}}}{I_0}\sim   
\varsigma
\frac{d}{\ell_{\text{ee}}}\frac{e\mathcal{E}d}{\epsilon_{\text{F}}}.
\end{equation}
Comparing Eqs.~\eqref{eq:dI-I-hydro} and \eqref{eq:dI-I-dis}
we see that in the hydrodynamic regime the elastic electron mean free path is effectively replaced by the channel width, $\ell_{\text{ei}}\to d$, and an additional large factor $d/\ell_{\text{ee}}\gg1$ arises from the dependence of the flow velocity on the channel width.  Thus, the relative nonreciprocal correction in the hydrodynamic regime is enhanced in comparison to the disorder-dominated regime by a large factor 
\begin{align}
    \label{eq:enhancement_factor}
    \frac{\delta I_{\text{hydro}}}{\delta I_{\text{dis}}} \sim \frac{d^2}{\ell_{\mathrm{ee}}\ell_{\mathrm{ei}}}.
\end{align}
In the presence of impurity scattering in the bulk, the hydrodynamic approach used above is valid provided 
$d$ does not exceed the Gurzhi length, $\ell_{\text{G}}=\sqrt{\ell_{\text{ei}}\ell_{\text{ee}}}$. For wide channels ($d>\ell_{\text{G}}$), electron flow is dominated by the scattering in the bulk of the device, leading to a transition from hydrodynamic Poiseuille flow to Ohmic behavior. As expected, at the crossover boundary between the two regimes, $d\sim \ell_{\text{G}}$, the results for nonreciprocity in Eqs.~\eqref{eq:dI-I-hydro} and \eqref{eq:dI-I-dis} match. 
 
We thank B. Spivak for useful discussions. 
The work of L. B. and A. V. A. was supported by the National Science Foundation Grant No. DMR-2424364, the Thouless Institute for Quantum Matter (TIQM), and the College of Arts and Sciences at the University of Washington. The work of A. L. was supported by NSF Grant No. DMR-2452658 and H. I. Romnes Faculty Fellowship provided by the University of Wisconsin-Madison Office of the Vice Chancellor for Research and Graduate Education with funding from the Wisconsin Alumni Research Foundation.

No data were created or analyzed in this study.

\bibliography{biblio}

\begin{thebibliography}{47}%
\makeatletter
\providecommand \@ifxundefined [1]{%
 \@ifx{#1\undefined}
}%
\providecommand \@ifnum [1]{%
 \ifnum #1\expandafter \@firstoftwo
 \else \expandafter \@secondoftwo
 \fi
}%
\providecommand \@ifx [1]{%
 \ifx #1\expandafter \@firstoftwo
 \else \expandafter \@secondoftwo
 \fi
}%
\providecommand \natexlab [1]{#1}%
\providecommand \enquote  [1]{``#1''}%
\providecommand \bibnamefont  [1]{#1}%
\providecommand \bibfnamefont [1]{#1}%
\providecommand \citenamefont [1]{#1}%
\providecommand \href@noop [0]{\@secondoftwo}%
\providecommand \href [0]{\begingroup \@sanitize@url \@href}%
\providecommand \@href[1]{\@@startlink{#1}\@@href}%
\providecommand \@@href[1]{\endgroup#1\@@endlink}%
\providecommand \@sanitize@url [0]{\catcode `\\12\catcode `\$12\catcode
  `\&12\catcode `\#12\catcode `\^12\catcode `\_12\catcode `\%12\relax}%
\providecommand \@@startlink[1]{}%
\providecommand \@@endlink[0]{}%
\providecommand \url  [0]{\begingroup\@sanitize@url \@url }%
\providecommand \@url [1]{\endgroup\@href {#1}{\urlprefix }}%
\providecommand \urlprefix  [0]{URL }%
\providecommand \Eprint [0]{\href }%
\providecommand \doibase [0]{http://dx.doi.org/}%
\providecommand \selectlanguage [0]{\@gobble}%
\providecommand \bibinfo  [0]{\@secondoftwo}%
\providecommand \bibfield  [0]{\@secondoftwo}%
\providecommand \translation [1]{[#1]}%
\providecommand \BibitemOpen [0]{}%
\providecommand \bibitemStop [0]{}%
\providecommand \bibitemNoStop [0]{.\EOS\space}%
\providecommand \EOS [0]{\spacefactor3000\relax}%
\providecommand \BibitemShut  [1]{\csname bibitem#1\endcsname}%
\let\auto@bib@innerbib\@empty
\bibitem [{\citenamefont {Onsager}(1931{\natexlab{a}})}]{Onsager-I}%
  \BibitemOpen
  \bibfield  {author} {\bibinfo {author} {\bibfnamefont {Lars}\ \bibnamefont
  {Onsager}},\ }\bibfield  {title} {\enquote {\bibinfo {title} {Reciprocal
  relations in irreversible processes. {I}.}}\ }\href {\doibase
  10.1103/PhysRev.37.405} {\bibfield  {journal} {\bibinfo  {journal} {Phys.
  Rev.}\ }\textbf {\bibinfo {volume} {37}},\ \bibinfo {pages} {405--426}
  (\bibinfo {year} {1931}{\natexlab{a}})}\BibitemShut {NoStop}%
\bibitem [{\citenamefont {Onsager}(1931{\natexlab{b}})}]{Onsager-II}%
  \BibitemOpen
  \bibfield  {author} {\bibinfo {author} {\bibfnamefont {Lars}\ \bibnamefont
  {Onsager}},\ }\bibfield  {title} {\enquote {\bibinfo {title} {Reciprocal
  relations in irreversible processes. {II}.}}\ }\href {\doibase
  10.1103/PhysRev.38.2265} {\bibfield  {journal} {\bibinfo  {journal} {Phys.
  Rev.}\ }\textbf {\bibinfo {volume} {38}},\ \bibinfo {pages} {2265--2279}
  (\bibinfo {year} {1931}{\natexlab{b}})}\BibitemShut {NoStop}%
\bibitem [{\citenamefont {Sturman}\ and\ \citenamefont
  {Fridkin}(1992)}]{Sturman:1992}%
  \BibitemOpen
  \bibfield  {author} {\bibinfo {author} {\bibfnamefont {B.}~\bibnamefont
  {Sturman}}\ and\ \bibinfo {author} {\bibfnamefont {V.}~\bibnamefont
  {Fridkin}},\ }\href@noop {} {\emph {\bibinfo {title} {Photovoltaic and
  Photorefractive effects in noncentrosymmetric materials}}},\ Ferroelectricity
  and related phenomena\ (\bibinfo  {publisher} {Taylor and Francis},\ \bibinfo
  {year} {1992})\BibitemShut {NoStop}%
\bibitem [{\citenamefont {Rikken}\ \emph {et~al.}(2001)\citenamefont {Rikken},
  \citenamefont {F\"olling},\ and\ \citenamefont {Wyder}}]{Rikken:PRL2001}%
  \BibitemOpen
  \bibfield  {author} {\bibinfo {author} {\bibfnamefont {G.~L. J.~A.}\
  \bibnamefont {Rikken}}, \bibinfo {author} {\bibfnamefont {J.}~\bibnamefont
  {F\"olling}}, \ and\ \bibinfo {author} {\bibfnamefont {P.}~\bibnamefont
  {Wyder}},\ }\bibfield  {title} {\enquote {\bibinfo {title} {Electrical
  magnetochiral anisotropy},}\ }\href {\doibase 10.1103/PhysRevLett.87.236602}
  {\bibfield  {journal} {\bibinfo  {journal} {Phys. Rev. Lett.}\ }\textbf
  {\bibinfo {volume} {87}},\ \bibinfo {pages} {236602} (\bibinfo {year}
  {2001})}\BibitemShut {NoStop}%
\bibitem [{\citenamefont {Ivchenko}\ and\ \citenamefont
  {Spivak}(2002)}]{Ivchenko:PRB2002}%
  \BibitemOpen
  \bibfield  {author} {\bibinfo {author} {\bibfnamefont {E.~L.}\ \bibnamefont
  {Ivchenko}}\ and\ \bibinfo {author} {\bibfnamefont {B.}~\bibnamefont
  {Spivak}},\ }\bibfield  {title} {\enquote {\bibinfo {title} {Chirality
  effects in carbon nanotubes},}\ }\href {\doibase 10.1103/PhysRevB.66.155404}
  {\bibfield  {journal} {\bibinfo  {journal} {Phys. Rev. B}\ }\textbf {\bibinfo
  {volume} {66}},\ \bibinfo {pages} {155404} (\bibinfo {year}
  {2002})}\BibitemShut {NoStop}%
\bibitem [{\citenamefont {Spivak}\ and\ \citenamefont
  {Zyuzin}(2004)}]{Spivak:PRL2004}%
  \BibitemOpen
  \bibfield  {author} {\bibinfo {author} {\bibfnamefont {B.}~\bibnamefont
  {Spivak}}\ and\ \bibinfo {author} {\bibfnamefont {A.}~\bibnamefont
  {Zyuzin}},\ }\bibfield  {title} {\enquote {\bibinfo {title} {Signature of the
  electron-electron interaction in the magnetic-field dependence of nonlinear
  {I}-{V} characteristics in mesoscopic systems},}\ }\href {\doibase
  10.1103/PhysRevLett.93.226801} {\bibfield  {journal} {\bibinfo  {journal}
  {Phys. Rev. Lett.}\ }\textbf {\bibinfo {volume} {93}},\ \bibinfo {pages}
  {226801} (\bibinfo {year} {2004})}\BibitemShut {NoStop}%
\bibitem [{\citenamefont {S\'anchez}\ and\ \citenamefont
  {B\"uttiker}(2004)}]{Sanchez:PRL2004}%
  \BibitemOpen
  \bibfield  {author} {\bibinfo {author} {\bibfnamefont {David}\ \bibnamefont
  {S\'anchez}}\ and\ \bibinfo {author} {\bibfnamefont {Markus}\ \bibnamefont
  {B\"uttiker}},\ }\bibfield  {title} {\enquote {\bibinfo {title}
  {Magnetic-field asymmetry of nonlinear mesoscopic transport},}\ }\href
  {\doibase 10.1103/PhysRevLett.93.106802} {\bibfield  {journal} {\bibinfo
  {journal} {Phys. Rev. Lett.}\ }\textbf {\bibinfo {volume} {93}},\ \bibinfo
  {pages} {106802} (\bibinfo {year} {2004})}\BibitemShut {NoStop}%
\bibitem [{\citenamefont {Polianski}\ and\ \citenamefont
  {B\"uttiker}(2006)}]{Polianski:PRL2006}%
  \BibitemOpen
  \bibfield  {author} {\bibinfo {author} {\bibfnamefont {M.~L.}\ \bibnamefont
  {Polianski}}\ and\ \bibinfo {author} {\bibfnamefont {M.}~\bibnamefont
  {B\"uttiker}},\ }\bibfield  {title} {\enquote {\bibinfo {title} {Mesoscopic
  fluctuations of nonlinear conductance of chaotic quantum dots},}\ }\href
  {\doibase 10.1103/PhysRevLett.96.156804} {\bibfield  {journal} {\bibinfo
  {journal} {Phys. Rev. Lett.}\ }\textbf {\bibinfo {volume} {96}},\ \bibinfo
  {pages} {156804} (\bibinfo {year} {2006})}\BibitemShut {NoStop}%
\bibitem [{\citenamefont {De~Martino}\ \emph {et~al.}(2006)\citenamefont
  {De~Martino}, \citenamefont {Egger},\ and\ \citenamefont
  {Tsvelik}}]{Tsvelik:PRL2006}%
  \BibitemOpen
  \bibfield  {author} {\bibinfo {author} {\bibfnamefont {A.}~\bibnamefont
  {De~Martino}}, \bibinfo {author} {\bibfnamefont {R.}~\bibnamefont {Egger}}, \
  and\ \bibinfo {author} {\bibfnamefont {A.~M.}\ \bibnamefont {Tsvelik}},\
  }\bibfield  {title} {\enquote {\bibinfo {title} {Nonlinear magnetotransport
  in interacting chiral nanotubes},}\ }\href {\doibase
  10.1103/PhysRevLett.97.076402} {\bibfield  {journal} {\bibinfo  {journal}
  {Phys. Rev. Lett.}\ }\textbf {\bibinfo {volume} {97}},\ \bibinfo {pages}
  {076402} (\bibinfo {year} {2006})}\BibitemShut {NoStop}%
\bibitem [{\citenamefont {Deyo}\ \emph {et~al.}(2006)\citenamefont {Deyo},
  \citenamefont {Spivak},\ and\ \citenamefont {Zyuzin}}]{Deyo:PRB2006}%
  \BibitemOpen
  \bibfield  {author} {\bibinfo {author} {\bibfnamefont {E.}~\bibnamefont
  {Deyo}}, \bibinfo {author} {\bibfnamefont {B.}~\bibnamefont {Spivak}}, \ and\
  \bibinfo {author} {\bibfnamefont {A.}~\bibnamefont {Zyuzin}},\ }\bibfield
  {title} {\enquote {\bibinfo {title} {Signature of the electron-electron
  interaction in the magnetic-field dependence of nonlinear {I}-{V}
  characteristics in mesoscopic conductors},}\ }\href {\doibase
  10.1103/PhysRevB.74.104205} {\bibfield  {journal} {\bibinfo  {journal} {Phys.
  Rev. B}\ }\textbf {\bibinfo {volume} {74}},\ \bibinfo {pages} {104205}
  (\bibinfo {year} {2006})}\BibitemShut {NoStop}%
\bibitem [{\citenamefont {L\"ofgren}\ \emph {et~al.}(2004)\citenamefont
  {L\"ofgren}, \citenamefont {Marlow}, \citenamefont {Shorubalko},
  \citenamefont {Taylor}, \citenamefont {Omling}, \citenamefont {Samuelson},\
  and\ \citenamefont {Linke}}]{Linke:PRL2004}%
  \BibitemOpen
  \bibfield  {author} {\bibinfo {author} {\bibfnamefont {A.}~\bibnamefont
  {L\"ofgren}}, \bibinfo {author} {\bibfnamefont {C.~A.}\ \bibnamefont
  {Marlow}}, \bibinfo {author} {\bibfnamefont {I.}~\bibnamefont {Shorubalko}},
  \bibinfo {author} {\bibfnamefont {R.~P.}\ \bibnamefont {Taylor}}, \bibinfo
  {author} {\bibfnamefont {P.}~\bibnamefont {Omling}}, \bibinfo {author}
  {\bibfnamefont {L.}~\bibnamefont {Samuelson}}, \ and\ \bibinfo {author}
  {\bibfnamefont {H.}~\bibnamefont {Linke}},\ }\bibfield  {title} {\enquote
  {\bibinfo {title} {Symmetry of two-terminal nonlinear electric conduction},}\
  }\href {\doibase 10.1103/PhysRevLett.92.046803} {\bibfield  {journal}
  {\bibinfo  {journal} {Phys. Rev. Lett.}\ }\textbf {\bibinfo {volume} {92}},\
  \bibinfo {pages} {046803} (\bibinfo {year} {2004})}\BibitemShut {NoStop}%
\bibitem [{\citenamefont {Wei}\ \emph {et~al.}(2005)\citenamefont {Wei},
  \citenamefont {Shimogawa}, \citenamefont {Wang}, \citenamefont {Radu},
  \citenamefont {Dormaier},\ and\ \citenamefont {Cobden}}]{Cobden:PRL2005}%
  \BibitemOpen
  \bibfield  {author} {\bibinfo {author} {\bibfnamefont {Jiang}\ \bibnamefont
  {Wei}}, \bibinfo {author} {\bibfnamefont {Michael}\ \bibnamefont
  {Shimogawa}}, \bibinfo {author} {\bibfnamefont {Zenghui}\ \bibnamefont
  {Wang}}, \bibinfo {author} {\bibfnamefont {Iuliana}\ \bibnamefont {Radu}},
  \bibinfo {author} {\bibfnamefont {Robert}\ \bibnamefont {Dormaier}}, \ and\
  \bibinfo {author} {\bibfnamefont {David~Henry}\ \bibnamefont {Cobden}},\
  }\bibfield  {title} {\enquote {\bibinfo {title} {Magnetic-field asymmetry of
  nonlinear transport in carbon nanotubes},}\ }\href {\doibase
  10.1103/PhysRevLett.95.256601} {\bibfield  {journal} {\bibinfo  {journal}
  {Phys. Rev. Lett.}\ }\textbf {\bibinfo {volume} {95}},\ \bibinfo {pages}
  {256601} (\bibinfo {year} {2005})}\BibitemShut {NoStop}%
\bibitem [{\citenamefont {Leturcq}\ \emph {et~al.}(2006)\citenamefont
  {Leturcq}, \citenamefont {S\'anchez}, \citenamefont {G\"otz}, \citenamefont
  {Ihn}, \citenamefont {Ensslin}, \citenamefont {Driscoll},\ and\ \citenamefont
  {Gossard}}]{Gossard:PRL2006}%
  \BibitemOpen
  \bibfield  {author} {\bibinfo {author} {\bibfnamefont {R.}~\bibnamefont
  {Leturcq}}, \bibinfo {author} {\bibfnamefont {D.}~\bibnamefont {S\'anchez}},
  \bibinfo {author} {\bibfnamefont {G.}~\bibnamefont {G\"otz}}, \bibinfo
  {author} {\bibfnamefont {T.}~\bibnamefont {Ihn}}, \bibinfo {author}
  {\bibfnamefont {K.}~\bibnamefont {Ensslin}}, \bibinfo {author} {\bibfnamefont
  {D.~C.}\ \bibnamefont {Driscoll}}, \ and\ \bibinfo {author} {\bibfnamefont
  {A.~C.}\ \bibnamefont {Gossard}},\ }\bibfield  {title} {\enquote {\bibinfo
  {title} {Magnetic field symmetry and phase rigidity of the nonlinear
  conductance in a ring},}\ }\href {\doibase 10.1103/PhysRevLett.96.126801}
  {\bibfield  {journal} {\bibinfo  {journal} {Phys. Rev. Lett.}\ }\textbf
  {\bibinfo {volume} {96}},\ \bibinfo {pages} {126801} (\bibinfo {year}
  {2006})}\BibitemShut {NoStop}%
\bibitem [{\citenamefont {Zumb\"uhl}\ \emph {et~al.}(2006)\citenamefont
  {Zumb\"uhl}, \citenamefont {Marcus}, \citenamefont {Hanson},\ and\
  \citenamefont {Gossard}}]{Marcus:PRL2006}%
  \BibitemOpen
  \bibfield  {author} {\bibinfo {author} {\bibfnamefont {D.~M.}\ \bibnamefont
  {Zumb\"uhl}}, \bibinfo {author} {\bibfnamefont {C.~M.}\ \bibnamefont
  {Marcus}}, \bibinfo {author} {\bibfnamefont {M.~P.}\ \bibnamefont {Hanson}},
  \ and\ \bibinfo {author} {\bibfnamefont {A.~C.}\ \bibnamefont {Gossard}},\
  }\bibfield  {title} {\enquote {\bibinfo {title} {Asymmetry of nonlinear
  transport and electron interactions in quantum dots},}\ }\href {\doibase
  10.1103/PhysRevLett.96.206802} {\bibfield  {journal} {\bibinfo  {journal}
  {Phys. Rev. Lett.}\ }\textbf {\bibinfo {volume} {96}},\ \bibinfo {pages}
  {206802} (\bibinfo {year} {2006})}\BibitemShut {NoStop}%
\bibitem [{\citenamefont {Angers}\ \emph {et~al.}(2007)\citenamefont {Angers},
  \citenamefont {Zakka-Bajjani}, \citenamefont {Deblock}, \citenamefont
  {Gu\'eron}, \citenamefont {Bouchiat}, \citenamefont {Cavanna}, \citenamefont
  {Gennser},\ and\ \citenamefont {Polianski}}]{Polianski:PRB2007}%
  \BibitemOpen
  \bibfield  {author} {\bibinfo {author} {\bibfnamefont {L.}~\bibnamefont
  {Angers}}, \bibinfo {author} {\bibfnamefont {E.}~\bibnamefont
  {Zakka-Bajjani}}, \bibinfo {author} {\bibfnamefont {R.}~\bibnamefont
  {Deblock}}, \bibinfo {author} {\bibfnamefont {S.}~\bibnamefont {Gu\'eron}},
  \bibinfo {author} {\bibfnamefont {H.}~\bibnamefont {Bouchiat}}, \bibinfo
  {author} {\bibfnamefont {A.}~\bibnamefont {Cavanna}}, \bibinfo {author}
  {\bibfnamefont {U.}~\bibnamefont {Gennser}}, \ and\ \bibinfo {author}
  {\bibfnamefont {M.}~\bibnamefont {Polianski}},\ }\bibfield  {title} {\enquote
  {\bibinfo {title} {Magnetic-field asymmetry of mesoscopic $\mathit{dc}$
  rectification in {A}haronov-{B}ohm rings},}\ }\href {\doibase
  10.1103/PhysRevB.75.115309} {\bibfield  {journal} {\bibinfo  {journal} {Phys.
  Rev. B}\ }\textbf {\bibinfo {volume} {75}},\ \bibinfo {pages} {115309}
  (\bibinfo {year} {2007})}\BibitemShut {NoStop}%
\bibitem [{\citenamefont {Nadeem}\ \emph {et~al.}(2023)\citenamefont {Nadeem},
  \citenamefont {Fuhrer},\ and\ \citenamefont {Wang}}]{Nadeem:2023}%
  \BibitemOpen
  \bibfield  {author} {\bibinfo {author} {\bibfnamefont {Muhammad}\
  \bibnamefont {Nadeem}}, \bibinfo {author} {\bibfnamefont {Michael~S.}\
  \bibnamefont {Fuhrer}}, \ and\ \bibinfo {author} {\bibfnamefont {Xiaolin}\
  \bibnamefont {Wang}},\ }\bibfield  {title} {\enquote {\bibinfo {title} {The
  superconducting diode effect},}\ }\href {\doibase 10.1038/s42254-023-00632-w}
  {\bibfield  {journal} {\bibinfo  {journal} {Nature Reviews Physics}\ }\textbf
  {\bibinfo {volume} {5}},\ \bibinfo {pages} {558--577} (\bibinfo {year}
  {2023})}\BibitemShut {NoStop}%
\bibitem [{\citenamefont {Liu}\ \emph {et~al.}(2024)\citenamefont {Liu},
  \citenamefont {Smith}, \citenamefont {Andreev},\ and\ \citenamefont
  {Spivak}}]{Liu:2024}%
  \BibitemOpen
  \bibfield  {author} {\bibinfo {author} {\bibfnamefont {T.}~\bibnamefont
  {Liu}}, \bibinfo {author} {\bibfnamefont {M.}~\bibnamefont {Smith}}, \bibinfo
  {author} {\bibfnamefont {A.~V.}\ \bibnamefont {Andreev}}, \ and\ \bibinfo
  {author} {\bibfnamefont {B.~Z.}\ \bibnamefont {Spivak}},\ }\bibfield  {title}
  {\enquote {\bibinfo {title} {Giant nonreciprocity of current-voltage
  characteristics of noncentrosymmetric superconductor--normal
  metal--superconductor junctions},}\ }\href {\doibase
  10.1103/PhysRevB.109.L020501} {\bibfield  {journal} {\bibinfo  {journal}
  {Phys. Rev. B}\ }\textbf {\bibinfo {volume} {109}},\ \bibinfo {pages}
  {L020501} (\bibinfo {year} {2024})}\BibitemShut {NoStop}%
\bibitem [{\citenamefont {Nagaosa}\ and\ \citenamefont
  {Yanase}(2024)}]{Yanase:2024}%
  \BibitemOpen
  \bibfield  {author} {\bibinfo {author} {\bibfnamefont {Naoto}\ \bibnamefont
  {Nagaosa}}\ and\ \bibinfo {author} {\bibfnamefont {Youichi}\ \bibnamefont
  {Yanase}},\ }\bibfield  {title} {\enquote {\bibinfo {title} {Nonreciprocal
  transport and optical phenomena in quantum materials},}\ }\href {\doibase
  https://doi.org/10.1146/annurev-conmatphys-032822-033734} {\bibfield
  {journal} {\bibinfo  {journal} {Annual Review of Condensed Matter Physics}\
  }\textbf {\bibinfo {volume} {15}},\ \bibinfo {pages} {63--83} (\bibinfo
  {year} {2024})}\BibitemShut {NoStop}%
\bibitem [{\citenamefont {Shliomis}(1967)}]{Shliomis:1967}%
  \BibitemOpen
  \bibfield  {author} {\bibinfo {author} {\bibfnamefont {MI}~\bibnamefont
  {Shliomis}},\ }\bibfield  {title} {\enquote {\bibinfo {title} {Hydrodynamics
  of a liquid with intrinsic rotation},}\ }\href@noop {} {\bibfield  {journal}
  {\bibinfo  {journal} {Soviet Journal of Experimental and Theoretical
  Physics}\ }\textbf {\bibinfo {volume} {24}},\ \bibinfo {pages} {173}
  (\bibinfo {year} {1967})}\BibitemShut {NoStop}%
\bibitem [{\citenamefont {Odenbach}(2002)}]{Odenbach:2002}%
  \BibitemOpen
  \bibfield  {author} {\bibinfo {author} {\bibfnamefont {Stefan}\ \bibnamefont
  {Odenbach}},\ }\href
  {https://search.library.wisc.edu/catalog/9910361582002121} {\emph {\bibinfo
  {title} {Magnetoviscous effects in ferrofluids}}}\ (\bibinfo  {publisher}
  {Berlin ; New York : Springer},\ \bibinfo {year} {2002})\BibitemShut
  {NoStop}%
\bibitem [{\citenamefont {Lifshitz}\ and\ \citenamefont
  {Pitaevskii}(1981)}]{LL-V10}%
  \BibitemOpen
  \bibfield  {author} {\bibinfo {author} {\bibfnamefont {E.~M.}\ \bibnamefont
  {Lifshitz}}\ and\ \bibinfo {author} {\bibfnamefont {L.~P.}\ \bibnamefont
  {Pitaevskii}},\ }\href@noop {} {\emph {\bibinfo {title} {{Course of
  theoretical physics. {V}ol. 10: Physical Kinetics}}}}\ (\bibinfo  {publisher}
  {Pergamon Press},\ \bibinfo {year} {1981})\BibitemShut {NoStop}%
\bibitem [{\citenamefont {Avron}(1998)}]{Avron:1998}%
  \BibitemOpen
  \bibfield  {author} {\bibinfo {author} {\bibfnamefont {J.~E.}\ \bibnamefont
  {Avron}},\ }\bibfield  {title} {\enquote {\bibinfo {title} {Odd viscosity},}\
  }\href {\doibase 10.1023/A:1023084404080} {\bibfield  {journal} {\bibinfo
  {journal} {Journal of Statistical Physics}\ }\textbf {\bibinfo {volume}
  {92}},\ \bibinfo {pages} {543--557} (\bibinfo {year} {1998})}\BibitemShut
  {NoStop}%
\bibitem [{\citenamefont {Haldane}(2009)}]{Haldane:2009}%
  \BibitemOpen
  \bibfield  {author} {\bibinfo {author} {\bibfnamefont {F.~D.~M.}\
  \bibnamefont {Haldane}},\ }\href {https://arxiv.org/abs/0906.1854} {\enquote
  {\bibinfo {title} {"{H}all viscosity" and intrinsic metric of incompressible
  fractional {H}all fluids},}\ } (\bibinfo {year} {2009}),\ \Eprint
  {http://arxiv.org/abs/0906.1854} {arXiv:0906.1854 [cond-mat.str-el]}
  \BibitemShut {NoStop}%
\bibitem [{\citenamefont {Bradlyn}\ \emph {et~al.}(2012)\citenamefont
  {Bradlyn}, \citenamefont {Goldstein},\ and\ \citenamefont
  {Read}}]{Read:2012}%
  \BibitemOpen
  \bibfield  {author} {\bibinfo {author} {\bibfnamefont {Barry}\ \bibnamefont
  {Bradlyn}}, \bibinfo {author} {\bibfnamefont {Moshe}\ \bibnamefont
  {Goldstein}}, \ and\ \bibinfo {author} {\bibfnamefont {N.}~\bibnamefont
  {Read}},\ }\bibfield  {title} {\enquote {\bibinfo {title} {Kubo formulas for
  viscosity: {H}all viscosity, {W}ard identities, and the relation with
  conductivity},}\ }\href {\doibase 10.1103/PhysRevB.86.245309} {\bibfield
  {journal} {\bibinfo  {journal} {Phys. Rev. B}\ }\textbf {\bibinfo {volume}
  {86}},\ \bibinfo {pages} {245309} (\bibinfo {year} {2012})}\BibitemShut
  {NoStop}%
\bibitem [{\citenamefont {Hoyos}\ and\ \citenamefont {Son}(2012)}]{Son:2012}%
  \BibitemOpen
  \bibfield  {author} {\bibinfo {author} {\bibfnamefont {Carlos}\ \bibnamefont
  {Hoyos}}\ and\ \bibinfo {author} {\bibfnamefont {Dam~Thanh}\ \bibnamefont
  {Son}},\ }\bibfield  {title} {\enquote {\bibinfo {title} {Hall viscosity and
  electromagnetic response},}\ }\href {\doibase 10.1103/PhysRevLett.108.066805}
  {\bibfield  {journal} {\bibinfo  {journal} {Phys. Rev. Lett.}\ }\textbf
  {\bibinfo {volume} {108}},\ \bibinfo {pages} {066805} (\bibinfo {year}
  {2012})}\BibitemShut {NoStop}%
\bibitem [{\citenamefont {Messica}\ \emph {et~al.}(2025)\citenamefont
  {Messica}, \citenamefont {Levchenko},\ and\ \citenamefont
  {Gutman}}]{Messica:2025}%
  \BibitemOpen
  \bibfield  {author} {\bibinfo {author} {\bibfnamefont {Yonatan}\ \bibnamefont
  {Messica}}, \bibinfo {author} {\bibfnamefont {Alex}\ \bibnamefont
  {Levchenko}}, \ and\ \bibinfo {author} {\bibfnamefont {Dmitri~B.}\
  \bibnamefont {Gutman}},\ }\href {https://arxiv.org/abs/2501.10890} {\enquote
  {\bibinfo {title} {Stokes flow in the electronic fluid with odd viscosity},}\
  } (\bibinfo {year} {2025}),\ \Eprint {http://arxiv.org/abs/2501.10890}
  {arXiv:2501.10890 [cond-mat.mes-hall]} \BibitemShut {NoStop}%
\bibitem [{\citenamefont {Banerjee}\ \emph {et~al.}(2017)\citenamefont
  {Banerjee}, \citenamefont {Souslov}, \citenamefont {Abanov},\ and\
  \citenamefont {Vitelli}}]{Banerjee:2017}%
  \BibitemOpen
  \bibfield  {author} {\bibinfo {author} {\bibfnamefont {Debarghya}\
  \bibnamefont {Banerjee}}, \bibinfo {author} {\bibfnamefont {Anton}\
  \bibnamefont {Souslov}}, \bibinfo {author} {\bibfnamefont {Alexander~G.}\
  \bibnamefont {Abanov}}, \ and\ \bibinfo {author} {\bibfnamefont {Vincenzo}\
  \bibnamefont {Vitelli}},\ }\bibfield  {title} {\enquote {\bibinfo {title}
  {Odd viscosity in chiral active fluids},}\ }\href {\doibase
  10.1038/s41467-017-01378-7} {\bibfield  {journal} {\bibinfo  {journal}
  {Nature Communications}\ }\textbf {\bibinfo {volume} {8}},\ \bibinfo {pages}
  {1573} (\bibinfo {year} {2017})}\BibitemShut {NoStop}%
\bibitem [{\citenamefont {Aggarwal}\ \emph {et~al.}(2023)\citenamefont
  {Aggarwal}, \citenamefont {Kirkinis},\ and\ \citenamefont {Olvera de~la
  Cruz}}]{Aggarwal2023}%
  \BibitemOpen
  \bibfield  {author} {\bibinfo {author} {\bibfnamefont {A.}~\bibnamefont
  {Aggarwal}}, \bibinfo {author} {\bibfnamefont {E.}~\bibnamefont {Kirkinis}},
  \ and\ \bibinfo {author} {\bibfnamefont {M.}~\bibnamefont {Olvera de~la
  Cruz}},\ }\bibfield  {title} {\enquote {\bibinfo {title} {Thermocapillary
  migrating odd viscous droplets},}\ }\href@noop {} {\bibfield  {journal}
  {\bibinfo  {journal} {{Physical Review Letters}}\ }\textbf {\bibinfo {volume}
  {131}},\ \bibinfo {pages} {198201} (\bibinfo {year} {2023})}\BibitemShut
  {NoStop}%
\bibitem [{\citenamefont {Kirkinis}\ and\ \citenamefont {Olvera de~la
  Cruz}(2023)}]{Kirkinis2023taylor}%
  \BibitemOpen
  \bibfield  {author} {\bibinfo {author} {\bibfnamefont {E.}~\bibnamefont
  {Kirkinis}}\ and\ \bibinfo {author} {\bibfnamefont {M.}~\bibnamefont {Olvera
  de~la Cruz}},\ }\bibfield  {title} {\enquote {\bibinfo {title} {Taylor
  columns and inertial-like waves in a three-dimensional odd viscous liquid},}\
  }\href@noop {} {\bibfield  {journal} {\bibinfo  {journal} {{Journal of Fluid
  Mechanics}}\ }\textbf {\bibinfo {volume} {973}},\ \bibinfo {pages} {A30}
  (\bibinfo {year} {2023})}\BibitemShut {NoStop}%
\bibitem [{\citenamefont {Kirkinis}\ and\ \citenamefont {Olvera de~la
  Cruz}(2024)}]{Kirkinis2024}%
  \BibitemOpen
  \bibfield  {author} {\bibinfo {author} {\bibfnamefont {E.}~\bibnamefont
  {Kirkinis}}\ and\ \bibinfo {author} {\bibfnamefont {M.}~\bibnamefont {Olvera
  de~la Cruz}},\ }\bibfield  {title} {\enquote {\bibinfo {title} {Evanescent
  and inertial-like waves in rigidly rotating odd viscous liquids},}\
  }\href@noop {} {\bibfield  {journal} {\bibinfo  {journal} {{Journal of Fluid
  Mechanics}}\ }\textbf {\bibinfo {volume} {996}},\ \bibinfo {pages} {A13}
  (\bibinfo {year} {2024})}\BibitemShut {NoStop}%
\bibitem [{\citenamefont {Narozhny}(2022)}]{Narozhny:2022}%
  \BibitemOpen
  \bibfield  {author} {\bibinfo {author} {\bibfnamefont {Boris~N.}\
  \bibnamefont {Narozhny}},\ }\bibfield  {title} {\enquote {\bibinfo {title}
  {Hydrodynamic approach to two-dimensional electron systems},}\ }\href
  {\doibase 10.1007/s40766-022-00036-z} {\bibfield  {journal} {\bibinfo
  {journal} {La Rivista del Nuovo Cimento}\ }\textbf {\bibinfo {volume} {45}},\
  \bibinfo {pages} {661--736} (\bibinfo {year} {2022})}\BibitemShut {NoStop}%
\bibitem [{\citenamefont {Fritz}\ and\ \citenamefont
  {Scaffidi}(2024)}]{Fritz:2024}%
  \BibitemOpen
  \bibfield  {author} {\bibinfo {author} {\bibfnamefont {L.}~\bibnamefont
  {Fritz}}\ and\ \bibinfo {author} {\bibfnamefont {T.}~\bibnamefont
  {Scaffidi}},\ }\bibfield  {title} {\enquote {\bibinfo {title} {Hydrodynamic
  electronic transport},}\ }\href {\doibase
  https://doi.org/10.1146/annurev-conmatphys-040521-042014} {\bibfield
  {journal} {\bibinfo  {journal} {Annual Review of Condensed Matter Physics}\
  }\textbf {\bibinfo {volume} {15}},\ \bibinfo {pages} {17--44} (\bibinfo
  {year} {2024})}\BibitemShut {NoStop}%
\bibitem [{\citenamefont {Zhou}\ \emph {et~al.}(2021)\citenamefont {Zhou},
  \citenamefont {Xie}, \citenamefont {Ghazaryan}, \citenamefont {Holder},
  \citenamefont {Ehrets}, \citenamefont {Spanton}, \citenamefont {Taniguchi},
  \citenamefont {Watanabe}, \citenamefont {Berg}, \citenamefont {Serbyn},\ and\
  \citenamefont {Young}}]{Zhou:2021}%
  \BibitemOpen
  \bibfield  {author} {\bibinfo {author} {\bibfnamefont {Haoxin}\ \bibnamefont
  {Zhou}}, \bibinfo {author} {\bibfnamefont {Tian}\ \bibnamefont {Xie}},
  \bibinfo {author} {\bibfnamefont {Areg}\ \bibnamefont {Ghazaryan}}, \bibinfo
  {author} {\bibfnamefont {Tobias}\ \bibnamefont {Holder}}, \bibinfo {author}
  {\bibfnamefont {James~R.}\ \bibnamefont {Ehrets}}, \bibinfo {author}
  {\bibfnamefont {Eric~M.}\ \bibnamefont {Spanton}}, \bibinfo {author}
  {\bibfnamefont {Takashi}\ \bibnamefont {Taniguchi}}, \bibinfo {author}
  {\bibfnamefont {Kenji}\ \bibnamefont {Watanabe}}, \bibinfo {author}
  {\bibfnamefont {Erez}\ \bibnamefont {Berg}}, \bibinfo {author} {\bibfnamefont
  {Maksym}\ \bibnamefont {Serbyn}}, \ and\ \bibinfo {author} {\bibfnamefont
  {Andrea~F.}\ \bibnamefont {Young}},\ }\bibfield  {title} {\enquote {\bibinfo
  {title} {Half- and quarter-metals in rhombohedral trilayer graphene},}\
  }\href {\doibase 10.1038/s41586-021-03938-w} {\bibfield  {journal} {\bibinfo
  {journal} {Nature}\ }\textbf {\bibinfo {volume} {598}},\ \bibinfo {pages}
  {429--433} (\bibinfo {year} {2021})}\BibitemShut {NoStop}%
\bibitem [{\citenamefont {Zhang}\ \emph {et~al.}(2024)\citenamefont {Zhang},
  \citenamefont {Lin}, \citenamefont {Chichinadze}, \citenamefont {Wang},
  \citenamefont {Watanabe}, \citenamefont {Taniguchi}, \citenamefont {Fu},\
  and\ \citenamefont {Li}}]{Zhang:2024}%
  \BibitemOpen
  \bibfield  {author} {\bibinfo {author} {\bibfnamefont {Naiyuan~James}\
  \bibnamefont {Zhang}}, \bibinfo {author} {\bibfnamefont {Jiang-Xiazi}\
  \bibnamefont {Lin}}, \bibinfo {author} {\bibfnamefont {Dmitry~V.}\
  \bibnamefont {Chichinadze}}, \bibinfo {author} {\bibfnamefont {Yibang}\
  \bibnamefont {Wang}}, \bibinfo {author} {\bibfnamefont {Kenji}\ \bibnamefont
  {Watanabe}}, \bibinfo {author} {\bibfnamefont {Takashi}\ \bibnamefont
  {Taniguchi}}, \bibinfo {author} {\bibfnamefont {Liang}\ \bibnamefont {Fu}}, \
  and\ \bibinfo {author} {\bibfnamefont {J.~I.~A.}\ \bibnamefont {Li}},\
  }\bibfield  {title} {\enquote {\bibinfo {title} {Angle-resolved transport
  non-reciprocity and spontaneous symmetry breaking in twisted trilayer
  graphene},}\ }\href {\doibase 10.1038/s41563-024-01809-z} {\bibfield
  {journal} {\bibinfo  {journal} {Nature Materials}\ }\textbf {\bibinfo
  {volume} {23}},\ \bibinfo {pages} {356--362} (\bibinfo {year}
  {2024})}\BibitemShut {NoStop}%
\bibitem [{\citenamefont {Lin}\ \emph {et~al.}(2023)\citenamefont {Lin},
  \citenamefont {Wang}, \citenamefont {Zhang}, \citenamefont {Watanabe},
  \citenamefont {Taniguchi}, \citenamefont {Fu},\ and\ \citenamefont
  {Li}}]{LeoLi:arxiv2023}%
  \BibitemOpen
  \bibfield  {author} {\bibinfo {author} {\bibfnamefont {Jiang-Xiazi}\
  \bibnamefont {Lin}}, \bibinfo {author} {\bibfnamefont {Yibang}\ \bibnamefont
  {Wang}}, \bibinfo {author} {\bibfnamefont {Naiyuan~J.}\ \bibnamefont
  {Zhang}}, \bibinfo {author} {\bibfnamefont {Kenji}\ \bibnamefont {Watanabe}},
  \bibinfo {author} {\bibfnamefont {Takashi}\ \bibnamefont {Taniguchi}},
  \bibinfo {author} {\bibfnamefont {Liang}\ \bibnamefont {Fu}}, \ and\ \bibinfo
  {author} {\bibfnamefont {J.~I.~A.}\ \bibnamefont {Li}},\ }\href@noop {}
  {\enquote {\bibinfo {title} {Spontaneous momentum polarization and diodicity
  in {B}ernal bilayer graphene},}\ } (\bibinfo {year} {2023}),\ \Eprint
  {http://arxiv.org/abs/2302.04261} {arXiv:2302.04261 [cond-mat.mes-hall]}
  \BibitemShut {NoStop}%
\bibitem [{\citenamefont {Chichinadze}\ \emph {et~al.}(2024)\citenamefont
  {Chichinadze}, \citenamefont {Zhang}, \citenamefont {Lin}, \citenamefont
  {Wang}, \citenamefont {Watanabe}, \citenamefont {Taniguchi}, \citenamefont
  {Vafek},\ and\ \citenamefont {Li}}]{Chichinadze:2024}%
  \BibitemOpen
  \bibfield  {author} {\bibinfo {author} {\bibfnamefont {Dmitry~V.}\
  \bibnamefont {Chichinadze}}, \bibinfo {author} {\bibfnamefont
  {Naiyuan~James}\ \bibnamefont {Zhang}}, \bibinfo {author} {\bibfnamefont
  {Jiang-Xiazi}\ \bibnamefont {Lin}}, \bibinfo {author} {\bibfnamefont
  {Xiaoyu}\ \bibnamefont {Wang}}, \bibinfo {author} {\bibfnamefont {Kenji}\
  \bibnamefont {Watanabe}}, \bibinfo {author} {\bibfnamefont {Takashi}\
  \bibnamefont {Taniguchi}}, \bibinfo {author} {\bibfnamefont {Oskar}\
  \bibnamefont {Vafek}}, \ and\ \bibinfo {author} {\bibfnamefont {J.~I.~A.}\
  \bibnamefont {Li}},\ }\href {https://arxiv.org/abs/2411.11156} {\enquote
  {\bibinfo {title} {Observation of giant nonlinear {H}all conductivity in
  {B}ernal bilayer graphene},}\ } (\bibinfo {year} {2024}),\ \Eprint
  {http://arxiv.org/abs/2411.11156} {arXiv:2411.11156 [cond-mat.mes-hall]}
  \BibitemShut {NoStop}%
\bibitem [{\citenamefont {Landau}\ and\ \citenamefont
  {Lifshitz}(1987)}]{LL-V6}%
  \BibitemOpen
  \bibfield  {author} {\bibinfo {author} {\bibfnamefont {L.~D.}\ \bibnamefont
  {Landau}}\ and\ \bibinfo {author} {\bibfnamefont {E.~M.}\ \bibnamefont
  {Lifshitz}},\ }\href@noop {} {\emph {\bibinfo {title} {Fluid Mechanics}}},\
  \bibinfo {edition} {2nd}\ ed.,\ \bibinfo {series} {Course of Theoretical
  Physics Series}, Vol.~\bibinfo {volume} {6}\ (\bibinfo  {publisher}
  {Butterworth-Heinemann, Oxford},\ \bibinfo {year} {1987})\BibitemShut
  {NoStop}%
\bibitem [{\citenamefont {Batchelor}(1967)}]{Batchelor1967}%
  \BibitemOpen
  \bibfield  {author} {\bibinfo {author} {\bibfnamefont {G.K.}\ \bibnamefont
  {Batchelor}},\ }\href@noop {} {\emph {\bibinfo {title} {An introduction to
  fluid dynamics}}}\ (\bibinfo  {publisher} {Cambridge University Press},\
  \bibinfo {year} {1967})\BibitemShut {NoStop}%
\bibitem [{\citenamefont {Andreev}\ and\ \citenamefont
  {Glazman}(2006)}]{Andreev:PRL2006}%
  \BibitemOpen
  \bibfield  {author} {\bibinfo {author} {\bibfnamefont {A.~V.}\ \bibnamefont
  {Andreev}}\ and\ \bibinfo {author} {\bibfnamefont {L.~I.}\ \bibnamefont
  {Glazman}},\ }\bibfield  {title} {\enquote {\bibinfo {title} {Nonlinear
  magnetoconductance of a classical ballistic system},}\ }\href {\doibase
  10.1103/PhysRevLett.97.266806} {\bibfield  {journal} {\bibinfo  {journal}
  {Phys. Rev. Lett.}\ }\textbf {\bibinfo {volume} {97}},\ \bibinfo {pages}
  {266806} (\bibinfo {year} {2006})}\BibitemShut {NoStop}%
\bibitem [{Note1()}]{Note1}%
  \BibitemOpen
  \bibinfo {note} {In the framework of the Boltzmann transport equation, these
  parameters can be obtained following the method of Enskog (see e.g. \protect
  \S 6 of Ref.~\cite {LL-V10}). It is important that for non-parabolic electron
  dispersion $\epsilon (\protect \bm {p})$, the locally-equilibrium
  distribution function in the presence of the hydrodynamic velocity $\protect
  \bm {v}$, $n_0 (\protect \bm {p}) = \left [e^{ (\epsilon (\protect \bm {p}) -
  \protect \bm {p}\cdot \protect \bm {v}-\mu )/T}+1 \right ]^{-1}$, does not
  amount to a shift of momentum and the chemical potential $\mu $, but alters
  the shape of the Fermi surface. The distortion of the Fermi surface caused by
  $\protect \bm {v}$ modifies all the kinetic coefficients of the electron
  liquid, including the viscosity. In electronic liquids with broken inversion
  and time-reversal symmetry, the modification of viscosity starts from the
  linear order in $\protect \bm {v}$.}\BibitemShut {Stop}%
\bibitem [{\citenamefont {Landau}\ and\ \citenamefont
  {Lifshitz}(1986)}]{LL-V7}%
  \BibitemOpen
  \bibfield  {author} {\bibinfo {author} {\bibfnamefont {L.~D.}\ \bibnamefont
  {Landau}}\ and\ \bibinfo {author} {\bibfnamefont {E.~M.}\ \bibnamefont
  {Lifshitz}},\ }\href@noop {} {\emph {\bibinfo {title} {Theory of
  Elasticity}}},\ \bibinfo {edition} {3rd}\ ed.,\ \bibinfo {series} {Course of
  Theoretical Physics Series}, Vol.~\bibinfo {volume} {7}\ (\bibinfo
  {publisher} {Butterworth-Heinemann, Oxford},\ \bibinfo {year}
  {1986})\BibitemShut {NoStop}%
\bibitem [{Note2()}]{Note2}%
  \BibitemOpen
  \bibinfo {note} {In systems lacking Galilean invariance, the total current
  density consists of the usual convective term, $enu$, and an additional
  contribution arising from the finite intrinsic conductivity~\cite {LAL:2022}.
  Assuming the particle density is sufficiently high, we neglect the
  latter.}\BibitemShut {Stop}%
\bibitem [{\citenamefont {Gurzhi}(1968)}]{Gurzhi:1968}%
  \BibitemOpen
  \bibfield  {author} {\bibinfo {author} {\bibfnamefont {R.~N.}\ \bibnamefont
  {Gurzhi}},\ }\bibfield  {title} {\enquote {\bibinfo {title} {Hydrodynamic
  effects in solids at low temperature},}\ }\href@noop {} {\bibfield  {journal}
  {\bibinfo  {journal} {Soviet Physics Uspekhi}\ }\textbf {\bibinfo {volume}
  {11}},\ \bibinfo {pages} {255} (\bibinfo {year} {1968})}\BibitemShut
  {NoStop}%
\bibitem [{\citenamefont {Krishna~Kumar}\ \emph {et~al.}(2017)\citenamefont
  {Krishna~Kumar}, \citenamefont {Bandurin}, \citenamefont {Pellegrino},
  \citenamefont {Cao}, \citenamefont {Principi}, \citenamefont {Guo},
  \citenamefont {Auton}, \citenamefont {Ben~Shalom}, \citenamefont
  {Ponomarenko}, \citenamefont {Falkovich}, \citenamefont {Watanabe},
  \citenamefont {Taniguchi}, \citenamefont {Grigorieva}, \citenamefont
  {Levitov}, \citenamefont {Polini},\ and\ \citenamefont {Geim}}]{Kumar:2017}%
  \BibitemOpen
  \bibfield  {author} {\bibinfo {author} {\bibfnamefont {R.}~\bibnamefont
  {Krishna~Kumar}}, \bibinfo {author} {\bibfnamefont {D.~A.}\ \bibnamefont
  {Bandurin}}, \bibinfo {author} {\bibfnamefont {F.~M.~D.}\ \bibnamefont
  {Pellegrino}}, \bibinfo {author} {\bibfnamefont {Y.}~\bibnamefont {Cao}},
  \bibinfo {author} {\bibfnamefont {A.}~\bibnamefont {Principi}}, \bibinfo
  {author} {\bibfnamefont {H.}~\bibnamefont {Guo}}, \bibinfo {author}
  {\bibfnamefont {G.~H.}\ \bibnamefont {Auton}}, \bibinfo {author}
  {\bibfnamefont {M.}~\bibnamefont {Ben~Shalom}}, \bibinfo {author}
  {\bibfnamefont {L.~A.}\ \bibnamefont {Ponomarenko}}, \bibinfo {author}
  {\bibfnamefont {G.}~\bibnamefont {Falkovich}}, \bibinfo {author}
  {\bibfnamefont {K.}~\bibnamefont {Watanabe}}, \bibinfo {author}
  {\bibfnamefont {T.}~\bibnamefont {Taniguchi}}, \bibinfo {author}
  {\bibfnamefont {I.~V.}\ \bibnamefont {Grigorieva}}, \bibinfo {author}
  {\bibfnamefont {L.~S.}\ \bibnamefont {Levitov}}, \bibinfo {author}
  {\bibfnamefont {M.}~\bibnamefont {Polini}}, \ and\ \bibinfo {author}
  {\bibfnamefont {A.~K.}\ \bibnamefont {Geim}},\ }\bibfield  {title} {\enquote
  {\bibinfo {title} {Superballistic flow of viscous electron fluid through
  graphene constrictions},}\ }\href {\doibase 10.1038/nphys4240} {\bibfield
  {journal} {\bibinfo  {journal} {Nature Physics}\ }\textbf {\bibinfo {volume}
  {13}},\ \bibinfo {pages} {1182--1185} (\bibinfo {year} {2017})}\BibitemShut
  {NoStop}%
\bibitem [{\citenamefont {Krebs}\ \emph {et~al.}(2023)\citenamefont {Krebs},
  \citenamefont {Behn}, \citenamefont {Li}, \citenamefont {Smith},
  \citenamefont {Watanabe}, \citenamefont {Taniguchi}, \citenamefont
  {Levchenko},\ and\ \citenamefont {Brar}}]{Brar:2023}%
  \BibitemOpen
  \bibfield  {author} {\bibinfo {author} {\bibfnamefont {Zachary~J.}\
  \bibnamefont {Krebs}}, \bibinfo {author} {\bibfnamefont {Wyatt~A.}\
  \bibnamefont {Behn}}, \bibinfo {author} {\bibfnamefont {Songci}\ \bibnamefont
  {Li}}, \bibinfo {author} {\bibfnamefont {Keenan~J.}\ \bibnamefont {Smith}},
  \bibinfo {author} {\bibfnamefont {Kenji}\ \bibnamefont {Watanabe}}, \bibinfo
  {author} {\bibfnamefont {Takashi}\ \bibnamefont {Taniguchi}}, \bibinfo
  {author} {\bibfnamefont {Alex}\ \bibnamefont {Levchenko}}, \ and\ \bibinfo
  {author} {\bibfnamefont {Victor~W.}\ \bibnamefont {Brar}},\ }\bibfield
  {title} {\enquote {\bibinfo {title} {Imaging the breaking of electrostatic
  dams in graphene for ballistic and viscous fluids},}\ }\href {\doibase
  10.1126/science.abm6073} {\bibfield  {journal} {\bibinfo  {journal}
  {Science}\ }\textbf {\bibinfo {volume} {379}},\ \bibinfo {pages} {671--676}
  (\bibinfo {year} {2023})}\BibitemShut {NoStop}%
\bibitem [{\citenamefont {Zeng}\ \emph {et~al.}(2024)\citenamefont {Zeng},
  \citenamefont {Guo}, \citenamefont {Ghosh}, \citenamefont {Watanabe},
  \citenamefont {Taniguchi}, \citenamefont {Levitov},\ and\ \citenamefont
  {Dean}}]{Zeng:2024}%
  \BibitemOpen
  \bibfield  {author} {\bibinfo {author} {\bibfnamefont {Yihang}\ \bibnamefont
  {Zeng}}, \bibinfo {author} {\bibfnamefont {Haoyu}\ \bibnamefont {Guo}},
  \bibinfo {author} {\bibfnamefont {Olivia~M.}\ \bibnamefont {Ghosh}}, \bibinfo
  {author} {\bibfnamefont {Kenji}\ \bibnamefont {Watanabe}}, \bibinfo {author}
  {\bibfnamefont {Takashi}\ \bibnamefont {Taniguchi}}, \bibinfo {author}
  {\bibfnamefont {Leonid~S.}\ \bibnamefont {Levitov}}, \ and\ \bibinfo {author}
  {\bibfnamefont {Cory~R.}\ \bibnamefont {Dean}},\ }\href
  {https://arxiv.org/abs/2407.05026} {\enquote {\bibinfo {title} {Quantitative
  measurement of viscosity in two-dimensional electron fluids},}\ } (\bibinfo
  {year} {2024}),\ \Eprint {http://arxiv.org/abs/2407.05026} {arXiv:2407.05026
  [cond-mat.mes-hall]} \BibitemShut {NoStop}%
\bibitem [{\citenamefont {Li}\ \emph {et~al.}(2022)\citenamefont {Li},
  \citenamefont {Andreev},\ and\ \citenamefont {Levchenko}}]{LAL:2022}%
  \BibitemOpen
  \bibfield  {author} {\bibinfo {author} {\bibfnamefont {Songci}\ \bibnamefont
  {Li}}, \bibinfo {author} {\bibfnamefont {A.~V.}\ \bibnamefont {Andreev}}, \
  and\ \bibinfo {author} {\bibfnamefont {Alex}\ \bibnamefont {Levchenko}},\
  }\bibfield  {title} {\enquote {\bibinfo {title} {Hydrodynamic electron
  transport in graphene {H}all-bar devices},}\ }\href {\doibase
  10.1103/PhysRevB.105.155307} {\bibfield  {journal} {\bibinfo  {journal}
  {Phys. Rev. B}\ }\textbf {\bibinfo {volume} {105}},\ \bibinfo {pages}
  {155307} (\bibinfo {year} {2022})}\BibitemShut {NoStop}%
\end{thebibliography}%

\end{document}